# Photoreactivity of condensed acetylene on Titan aerosols analogues

Benjamin Fleury[1], Murthy S. Gudipati[1,*], Isabelle Couturier-Tamburelli[2], Nathalie Carrasco[3,4]

[1] Science Division, Jet Propulsion Laboratory, California Institute of Technology, 4800 Oak Grove Drive, Pasadena, California 91109, USA

[2]Aix-Marseille Université, CNRS, PIIM, UMR 7345, 13013 Marseille, France

[3]Université Versailles St-Quentin, Sorbonne Universités, UPMC Univ. Paris 06, CNRS/INSU, LATMOS-IPSL, 11 Blvd. d'Alembert, 78280 Guyancourt, France

[4]Institut Universitaire de France, 103 Bvd St-Michel, 75005 Paris, France

*Corresponding author, e-mail: murthy.gudipati@jpl.nasa.gov

Pages: 28

Figures: 10

Tables: 1

# Abstract

Volatile organic molecules formed by photochemistry in the upper atmosphere of Titan can undergo condensation as pure ices in the stratosphere and the troposphere as well as condense as ice layers onto the organic aerosols that are visible as the haze layers of Titan. As solar photons penetrate through Titan's atmosphere, shorter-wavelength photons are attenuated and longer-wavelength photons make it into the lower altitudes, where aerosols become abundant. We conducted an experimental study to evaluate the long wavelength ($\lambda > 300$ nm) photo-reactivity of these ices accreted on the Titan aerosol-analogs (also known as tholins) made in the laboratory. We have focused on acetylene, the third most abundant hydrocarbon in Titan's atmosphere (after $CH_4$ and $C_2H_6$). Further, acetylene is the most abundant unsaturated hydrocarbon in Titan's atmosphere. Our results indicate that the aerosols can act as activation centers to drive the photoreactivity of acetylene with the aerosols at the accretion interface at wavelengths where acetylene-ice alone does not show photoreactivity. We found that along with photochemistry, photodesorption plays an important role. We observed that about 15% of the initial acetylene is photodesorbed, with a photodesorption rate of $(2.1 \pm 0.2) \times 10^{-6}$ molecules·photon$^{-1}$ at 355 nm. This photodesorption is wavelength-dependent, confirming that it is mediated by the UV absorption of the aerosol analogues, similar to photochemistry. We conclude that the UV-Vis properties of aerosols would determine how they evolve further in Titan's atmosphere and on the surface through photochemical alterations involving longer-wavelength photons. Stronger extinction coefficients in the longer wavelength UV-Vis regions (>300 nm) of the aerosol give higher efficiencies of photodesorption of accreted volatiles as well as photochemical incorporation of unsaturated condensates (such as acetylene) into the aerosols.

# 1. Introduction

Titan, the largest satellite of Saturn, has a dense atmosphere mainly made of molecular nitrogen and methane, the latter of which varies by 1% to 5% as a function of the altitude (Cui et al., 2009; Flasar et al., 2005; Hörst, 2017; Koskinen et al., 2011; Niemann et al., 2010). Dissociation and ionization of these two species by the vacuum ultraviolet (VUV, wavelengths below 200 nm) solar photons and the energetic particles of Saturn's magnetosphere initiate complex organic chemistry leading to the formation of volatile organic compounds and solid organic aerosols in Titan's upper atmosphere. Aerosol formation is initiated at about 1000 km, in the ionosphere of Titan (Lavvas et al., 2013; Liang et al., 2007; Waite et al., 2007). At lower altitudes, as the temperature decreases with altitude from the lower stratosphere (~150 km) towards the tropopause (Fulchignoni et al., 2005), many volatiles can condense to form ice particles or ice accreted on the aerosols (Anderson and Samuelson, 2011; Anderson et al., 2010; Anderson et al., 2016; Barth, 2017; Barth and Toon, 2003; de Kok et al., 2008; Frère et al., 1990; Lavvas et al., 2011a; Mayo and Samuelson, 2005; Raulin and Owen, 2002; Sagan and Reid Thompson, 1984; Samuelson and A. Mayo, 1991; Samuelson et al., 1997). These ice particles and aerosols precipitate and deposit onto the surface, where they should have formed an organic layer over a geological time-scale according to a study performed using a 1D photochemical model (Vuitton et al., 2008). As the condensates and aerosols increase in number at these altitudes, solar photons with shorter-wavelengths (higher energy) are attenuated, leaving only the longer-wavelength (>300 nm) photons to penetrate through. Understanding the photochemical transformations in these ices and aerosols caused by the solar photons that are available in these environments is the principal goal of our work.

There are a number of observations realized with Voyager IRIS and more recently Cassini CIRS and VIMS, which report the detection of ice particles in the stratosphere of Titan. Evidence for dicyanoacetylene ($C_4N_2$) ice was found in IRIS data (Khanna et al., 1987; Samuelson et al., 1997) and CIRS data (Anderson et al., 2016). The presence of cyanoacetylene ($HC_3N$) was also suggested by IRIS data and laboratory spectra (Coustenis et al., 1999; Khanna, 2005b) before being confirmed by CIRS observations (Anderson et al., 2010). Hydrogencyanide (HCN) ice has been tentatively identified using CIRS observations (Anderson and Samuelson, 2011; Samuelson et al., 2007) before being confirmed with VIMS (de Kok et al., 2014). The presence of other ices such as $C_2H_5CN$, $C_2H_6$, and $C_2H_2$ has also been proposed based on IRIS data and laboratory spectra (Coustenis et al., 1999; Khanna, 2005a; Khanna, 2005b) but they have not been confirmed.

Presence of ices is also reported at the surface of Titan based on Cassini and ground-based observations. Presence of water ice has been inferred from observation done at NASA's Infrared Telescope Facility (IRTF) (Griffith et al., 2003), using Cassini VIMS data (Griffith et al., 2012; Hayne et al., 2014; McCord et al., 2006) and the DISR instrument onboard the Huygens lander (Rannou et al., 2016; Tomasko et al., 2005). Other condensed ices have been suggested based on Cassini VIMS observations, but their identifications remain uncertain. These ices include $CO_2$ (McCord et al., 2008), $C_6H_6$, $HC_3N$ (Clark et al., 2010) and more recently $C_2H_2$ (Singh et al., 2016).

Acetylene is the third most abundant hydrocarbon present in Titan's atmosphere after $CH_4$ and $C_2H_6$ and condensed acetylene should be particularly abundant on Titan. Further, acetylene is the most abundant unsaturated hydrocarbon in Titan's atmosphere, making it a more plausible candidate for photochemical transformations in the condensed-phase. The $C_2H_2$ mixing ratio ranges from 2 to 6 $ppm_v$ as a function of latitude as measured in the stratosphere by Cassini CIRS (Vinatier et al., 2010). Formation of stratospheric $C_2H_2$ ice particles is predicted by modeling studies with a condensation near 70 km (Anderson and Samuelson, 2011; Barth, 2017; Lavvas et al., 2011a; Raulin and Owen, 2002; Sagan and Reid Thompson, 1984), but not yet detected in Titan's atmosphere. However, evaporated acetylene has been detected from Titan's surface by the GC-MS instrument of the Huygens probe after its landing (Niemann et al., 2010), suggesting that acetylene ice is present on the surface and also recently confirmed by the Cassini VIMS observation of Titan's surface (Singh et al., 2016).

Because solar photons at longer-wavelengths reach lower altitudes, condensed molecules could undergo further processes driven by those long wavelength photons ($\lambda > 300$ nm), particularly at and below the lower stratosphere (<150 km) where temperatures also drop below 150 K, all the way to the surface of Titan (Lavvas et al., 2008). Interaction of these photons with the aerosols and ices could lead to two different processes: (a) photodesorption of the volatile condensate molecules through direct or indirect mechanisms (Thrower et al., 2008), (b) long-wavelength photons could induce solid-state photochemistry leading to the formation of covalent bonds and new condensed volatiles organic compounds (Anderson et al., 2016) as well as solid organics like aerosols as demonstrated recently in laboratory experiments (Couturier-Tamburelli et al., 2014; Couturier-Tamburelli et al., 2015; Gudipati et al., 2013). One or several layers of ices could cover Titan's aerosols in the atmosphere or at the surface (Anderson et al., 2016; Raulin and Owen, 2002). Hence, photo-induced processes such photochemistry or photodesorption could alter the mixing ratios of volatiles and condensates in the atmosphere of Titan.

In this work, we have used laboratory experiments to study the photo-induced processes involving condensed acetylene in Titan's atmosphere, for photons at $\lambda > 300$ nm that reach the lower-stratosphere (<150 km) of Titan. We investigated the interaction between Titan's aerosols accreted with acetylene ice-films using laboratory analogs of Titan aerosols, called "tholins", that are made through cold-plasma discharge of $N_2$ and $CH_4$ gas mixture (Szopa et al., 2006). We use "tholins" to generally describe Titan's organic aerosol analogs made in the laboratory environment using plasma discharge, in agreement with literature usage (Cable et al., 2012).

## 2. Experimental methods and protocols

### 2.1. Tholin samples production

Tholin samples used in this study were produced with the PAMPRE experimental set-up, which simulates aerosol production in Titan's atmosphere (Sciamma-O'Brien et al., 2010; Szopa et al., 2006). A $N_2$-$CH_4$ gas mixture was subjected to radio-frequency discharge initiating a cold plasma at low pressures (0.9 mbar) and room temperature. Tholins accumulate on sapphire substrates (25 mm of diameter, 2 mm of thickness) disposed on the grounded electrode and form thin-film deposits. The percentage of methane used to produce the tholin samples is known to influence their compositions (Gautier et al., 2012) and their optical properties (Mahjoub et al., 2012). In order to study the effect of the tholins composition and optical properties on the acetylene photo-reactivity, we have produced tholins with 1% and 5% of $CH_4$ diluted in $N_2$, in agreement with the methane concentration profile in the atmosphere of Titan (Hörst, 2017): We call the tholins produced by using the 1% $CH_4$ as $T_1$-tholin from now onwards. Accordingly, $T_5$-tholin was obtained using the 5% methane. For a 2-hour plasma duration, the thickness of the tholin films varied between 630 nm and 910 nm based on the initial methane percentage (Mahjoub et al., 2012).

## 2.2. Experimental setup

The photoreactivity of solid acetylene was studied at JPL with an experimental setup that we have named the "TOAST" (Titan Organic Aerosol SpecTroscopy) setup (Couturier-Tamburelli et al., 2014; Gudipati et al., 2013). A schematic diagram of the TOAST setup is presented in Figure 1. It is composed of a vacuum chamber connected to a turbomolecular pump assuring an ultrahigh vacuum of $3 \times 10^{-8}$ mbar as measured by an inverted magnetron gauge at room temperature.

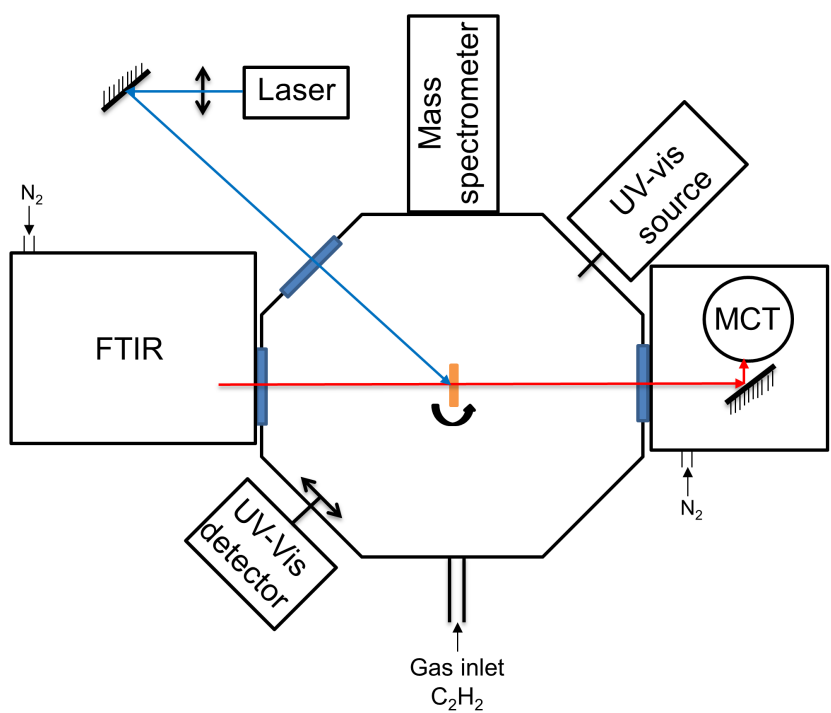

**Figure 1: Scheme of the TOAST experimental setup at JPL to simulate and study Titan's atmospheric and surface photochemistry.**

Acetylene is highly reactive and easily polymerizes at ambient conditions when stored for a long time. For this reason, we used a gas cylinder of 10% acetylene diluted in helium (Airgas) at 13 bar total pressure, where acetylene is stabilized over a long period of time (for years), and this gas mixture was deposited on a sapphire window (or a sapphire window coated with a tholin film), cooled down to 50 K using an ARS closed-cycle helium cryostat. Background pressure during the deposition in the vacuum chamber was kept at $1\times10^{-6}$ mbar and the typical duration of deposition was for 20 min. At this temperature, helium does not stick to the sapphire window or the tholin film and it is pumped off, leaving pure acetylene sticking to the sample. Control experiments with pure acetylene ice indicated that acetylene films start significantly sublimating at 70 K and completely sublime at 80 K from the sapphire window under the typically used background vacuum of ~$5\times10^{-9}$ mbar. We chose to keep the temperature at 50 K for all the experiments, which is lower than the temperature in Titan's atmosphere where acetylene would condense (Fulchignoni et al., 2005), because the pressure at which the experiments were performed (i.e. ~$5\times10^{-9}$ mbar) which was, by necessity, lower than that found in Titan's stratosphere, would cause thermal sublimation of acetylene. Control experiments confirmed that there was no detectable depletion of infrared absorption bands of acetylene ice over several hours at 50 K. It has been shown that during the warm up of amorphous $C_2H_2$ deposited at <20 K, amorphous-crystalline phase transition occurred at 45-50 K (Hudson et al., 2014; Knez et al., 2012). Therefore, under our experimental conditions and according to the position and shape of the $\upsilon_3$ band of $C_2H_2$ presented in Figure 2, we assume that acetylene ice is predominantly in a crystalline form.

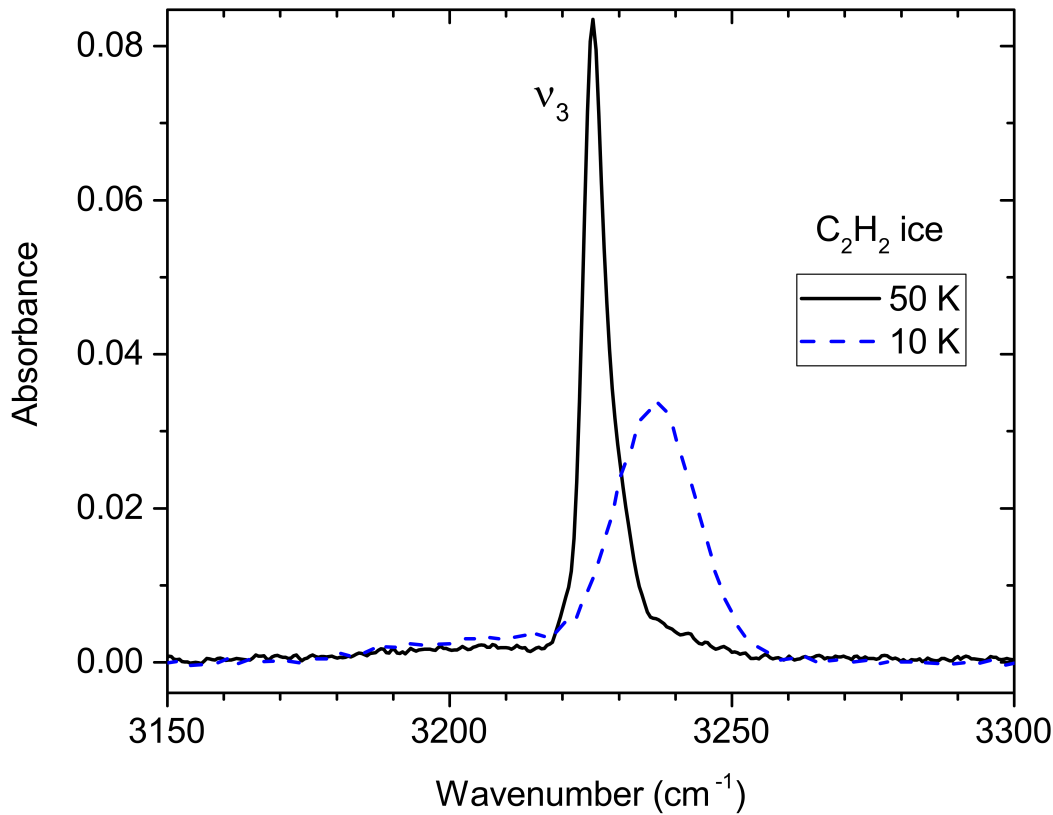

**Figure 2: $C_2H_2$ absorption band at 3225 cm$^{-1}$ ($\upsilon_3$) after 20 min deposition at 50 K (crystalline) vs. at 10 K (amorphous).**

Subsequent to the preparation of the acetylene film on sapphire or sapphire with tholin film, this sample was irradiated at different wavelengths to simulate aerosol interaction with solar photons reaching the lower atmosphere of Titan (< 150 km): 355 nm, 410 nm and 450 nm. These wavelengths were generated by a Nd:YAG laser (Quantel) with third harmonic generation (355 nm) and by tuning an optical parametric oscillator (OPO) (Opotek Opolette-UV) that provides continuous wavelengths between 210 nm and 700 nm. In order to avoid multi-photon processes and to irradiate the entire sample's surface, the initial laser beam of 3 mm of diameter was expanded to 20 mm. For 355 nm, the laser beam, which is circular, was expanded using a plano-convex lens (75 mm focal length). For the 410 nm and 450 nm, the OPO-laser beam that is highly elliptical (2 mm x 5 mm at 10 cm from the laser exit), was expanded using two plano-convex cylindrical lenses (100 mm and 200 mm focal lengths). Photon flux was measured at the entry of the vacuum chamber, after beam expansion. The calculated photon fluxes reaching the sample for each wavelength are given in Table 1.

**Table 1: Photon fluxes reaching the sample at different selected laser wavelength.**

| Laser wavelength (nm) | 355 | 410 | 450 |
|---|---|---|---|
| Photon flux (photons·cm$^{-2}$·s$^{-1}$) | $(1.80 \pm 0.05) \times 10^{17}$ | $(2.50 \pm 0.05) \times 10^{16}$ | $(2.70 \pm 0.05) \times 10^{16}$ |

Infrared (FTIR) and UV-Vis spectroscopy in transmission mode were used to monitor the evolution of acetylene/tholin samples during laser irradiation. Photodesorption of acetylene was monitored using *in situ* mass spectrometry.

### 2.3. Infrared absorption spectroscopy

Single-beam infrared spectra of the samples were measured with a Thermo Scientific Nicolet 6700 Fourier Transform Infrared Spectrometer (FTIR), which are processed to obtain absorption spectra by choosing appropriate spectra as reference and sample. As shown in Figure 1, the beam enters in the vacuum chamber through a ZnSe window, goes out through a second ZnSe window and is focused on to a Mercury Cadmium Telluride (MCT) detector cooled to 77 K by liquid nitrogen. Infrared spectra were recorded between 1600 cm$^{-1}$ (sapphire window absorption limit) and 4500 cm$^{-1}$ at 1 cm$^{-1}$ resolution by averaging 200 scans.

### 2.4. UV-visible absorption spectroscopy

The UV/visible spectra were measured at a 45° incidence angle to the sample window (Figure 1), using an Ocean Optics DH-2000 fiber coupled deuterium-halogen source. The output light was focused by a lens and collected by an Ocean Optics USB4000 fiber coupled spectrograph. Spectra were acquired in the 220-1100 nm range by averaging 500 scans recorded with an integration time of 20 ms per scan.

### 2.5. Mass spectrometry analysis of the gas-phase molecules

*In situ* measurements of photon-induced desorption of acetylene were carried out using a Stanford Research System RGA200 quadrupole mass spectrometer (QMS) equipped with electron-multiplier to increase sensitivity, which was directly connected to the main vacuum chamber with the sample. QMS ionization was achieved through electron impact at 70 eV. It covers 1 to 200 *m/z* mass range with a resolution of 100 at *m/z* 100 (*m/Δm*). The QMS measures overall equilibrated content of rest gases in the vacuum chamber. During the thermal and photodesorption experiments, the turbomolecular pump continuously removed gases in the vacuum chamber. Thus, the detected ion currents are quantitative and directly comparable between experiments as discussed below.

## 2.6. Quantification of $C_2H_2$ Ice Desorption

The total ion current measured at *m/z* 26 ($C_2H_2^+$) during the irradiation experiments is proportional to the number of acetylene molecules desorbed. In order to calculate the fraction of $C_2H_2$ molecules photodesorbed relative to the initial number of molecules $N_0$ that were deposited, we have first measured the total ion current at *m/z* 26 corresponding to the thermal desorption of the entire acetylene deposited. The same amount of acetylene as used in the irradiation experiments has been desorbed using temperature-programmed desorption (TPD) at 1 K·min$^{-1}$. The TPD curve measured for $C_2H_2$ is presented in Figure 3. The amount of acetylene deposited for each experiment is controlled using the absorbance of the 3225 cm$^{-1}$ band, which is proportional to the thickness of the ice deposited. The thickness of the deposited acetylene ice has been calculated using the Eq. (1):

$$h = \frac{\int \tau_v dv}{A \times \rho_N}, \qquad (1)$$

where $\int \tau_v dv$ is the integral of the band in optical depth, which can be derived from the absorbance value. A is the band-strength (cm.molecule$^{-1}$) and $\rho_N$ the number density (molecules.cm$^{-3}$). For our calculation, we used a band strength value of 3.58×10$^{-17}$ cm·molecule$^{-1}$ for the 3225 cm$^{-1}$ band calculated by Hudson et al. (2014) for the crystalline acetylene at 70 K. We used a number density of 1.76×10$^{22}$ molecules·cm$^{-3}$ derived from the mass density, assuming a value of 0.76 g·cm$^{-3}$ (Hudson et al., 2014; McMullan et al., 1992), Avogadro's number and the molar mass of acetylene (26.04 g·mol$^{-1}$). We calculated an average thickness of the acetylene ice film to be between 15 nm and 20 nm, based on minor variations in the deposition conditions that reflect in acetylene absorbance at 3225 cm$^{-1}$.

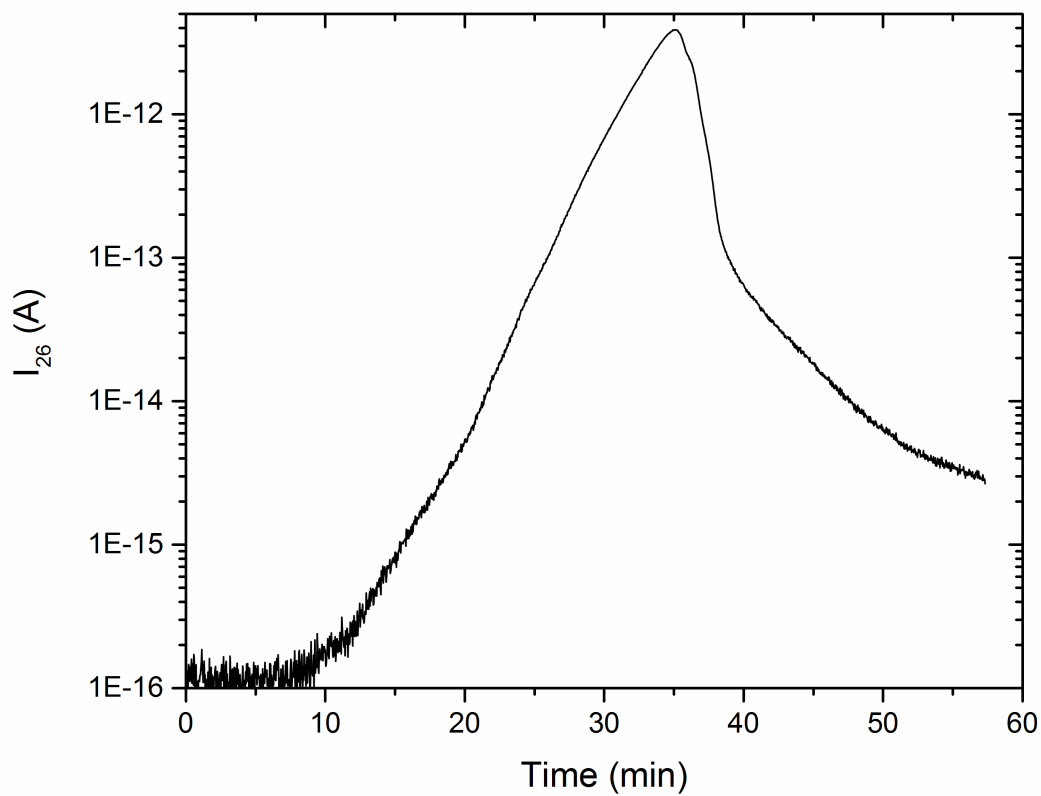

**Figure 3: Evolution of *m/z* = 26 (C$_2$H$_2^+$) intensity as function of time during the temperature-programmed desorption (TPD) of a C$_2$H$_2$ ice film prepared at 50 K.**

The integrated signal at *m/z* 26 measured during the TPD is proportional to the amount of desorbed C$_2$H$_2$ molecules N$_0$, which is used to calculate the percentage of acetylene photodesorbed during the laser irradiation. Compared to TPD, which is a continuous process, laser irradiation at 20 Hz repetition rate is a pulsed excitation (5 ns laser pulse width), which is expected to generate photodesorption within a shorter period of time in microseconds range, until the next laser pulse comes 50 milliseconds later. In order to ensure that such pulsed ejection of acetylene into the vacuum chamber is measured accurately, we integrated the ion-current at each mass for 400 milliseconds, averaging 8 laser pulses. Four mass peaks were monitored through this process, namely: H$_2$O$^+$ (*m/z* 18), C$_2$H$^+$ (*m/z* 25), C$_2$H$_2^+$ (*m/z* 26) and O$_2^+$ (*m/z* 32). Thus, our laser photodesorption data can be quantitatively compared with TPD data for acetylene.

# 3. Results

## 3.1. Acetylene photo-reactivity experiments

### 3.1.1. Pure $C_2H_2$-ice and $C_2H_2$ on $T_1$-tholin

We began our study with the photo-reactivity of acetylene deposited on a sapphire window or a $T_1$-tholin sample and irradiation with a laser at 355 nm. We have chosen to first focus on $T_1$-tholin, which has higher nitrogen incorporation than the $T_5$-tholin. The evolution of the acetylene amount on the sample was monitored using its infrared band at 3225 cm$^{-1}$, attributed to the $\upsilon_3$ vibrational stretching mode of the C-H bond (Hudson et al., 2014).

Figure 4 presents the evolution of the differential absorbance after each irradiation for pure $C_2H_2$ in the left panel and for $C_2H_2$ deposited on a $T_1$-tholin sample in the right panel. The differential absorbance is the absorbance after a time *"t"* of irradiation $A_t$ less the initial absorbance $A_{t=0}$. Figure 4 shows a decrease of the acetylene absorbance after irradiation at 355 nm, indicating depletion of acetylene in the sample. This photo-depletion is enhanced by an order of magnitude when acetylene was deposited on a $T_1$-tholin compared to the pure $C_2H_2$-ice on the sapphire substrate experiment.

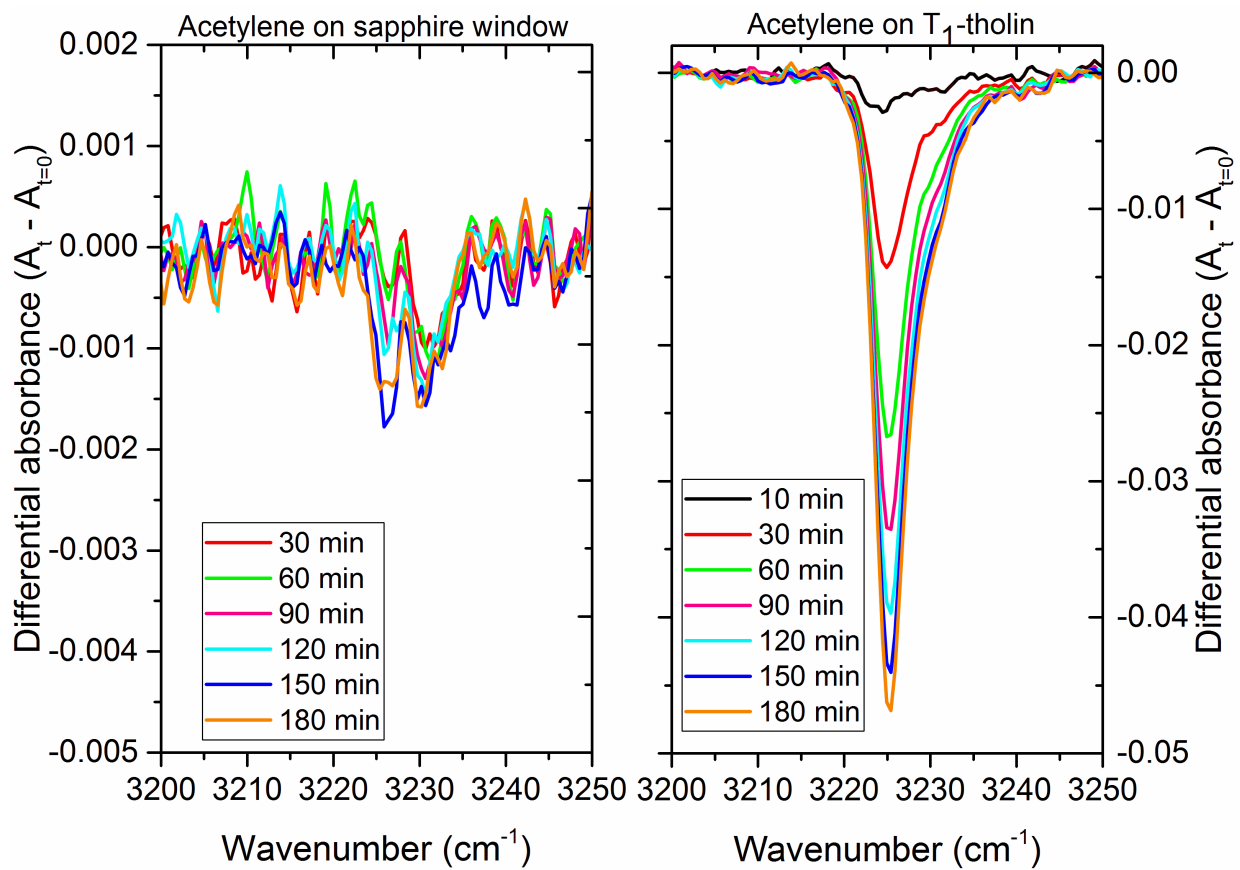

**Figure 4:** Evolution of the differential absorbance ($A_t - A_{t=0}$) of $C_2H_2$ as a function of the irradiation duration at 355 nm. The left panel presents results for acetylene coated on blank sapphire window. The right panel presents results for acetylene coated on a $T_1$-tholin sample.

Two possible photo-induced processes can be considered to explain the observed acetylene's depletion, namely, photodesorption and photochemistry. The photodesorption of $C_2H_2$ was monitored by *in situ* mass spectrometry using the time-tracking of $C_2H_2^+$ at *m/z* 26 in the residual gas phase during the irradiation. Figure 5 presents the evolution of the ion current measured by the mass spectrometer for *m/z* 26 as function of time during the irradiation at 355 nm of $C_2H_2$ on a $T_1$-tholin sample. When the laser was turned on, we observed an increase of the signal by two orders of magnitude. The intensity decreased slowly during the irradiation and then returned to its initial value when the laser was turned off. This increase of the signal measured by the QMS during the irradiation highlights the photodesorption of acetylene during the irradiation at 355 nm.

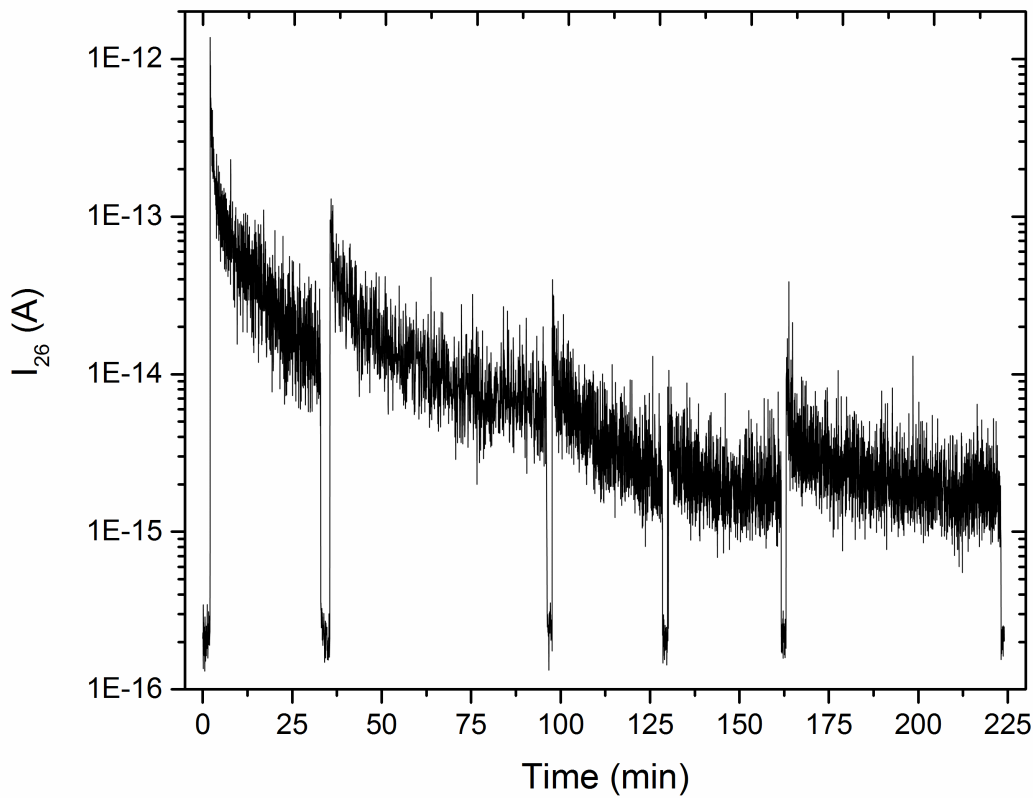

**Figure 5:** Evolution of $m/z = 26$ ($C_2H_2^+$) intensity as function of the time for the different irradiations step at 355 nm of $C_2H_2$ deposited on a $T_1$-tholin sample. Origin of time is set as the moment when the laser was turned on for the first time. Interruptions are the time-windows when IR were measured.

We have measured the integrated signal for each irradiation and calculated the fraction of the initial amount of acetylene photodesorbed for each irradiation (presented in Section 2.6). Similarly, we have calculated for each irradiation the amount of acetylene consumed, using the integrated band area at 3225 cm$^{-1}$ in the IR absorption spectrum. With this procedure, we can compare the evolution of the acetylene amount photodesorbed, measured with the QMS and remaining as ice film, measured using IR spectroscopy. Figure 6 presents the evolution of the amount of acetylene as a function of the irradiation time as measured by IR spectroscopy as well as the evolution of gas-phase acetylene quantity taking into account the acetylene desorbed as measured by QMS. Please note that due to negligible photodesorption of pure acetylene ice from the sapphire window, we have only quantified photodesorption for the experiment where acetylene was deposited on the $T_1$-tholin sample.

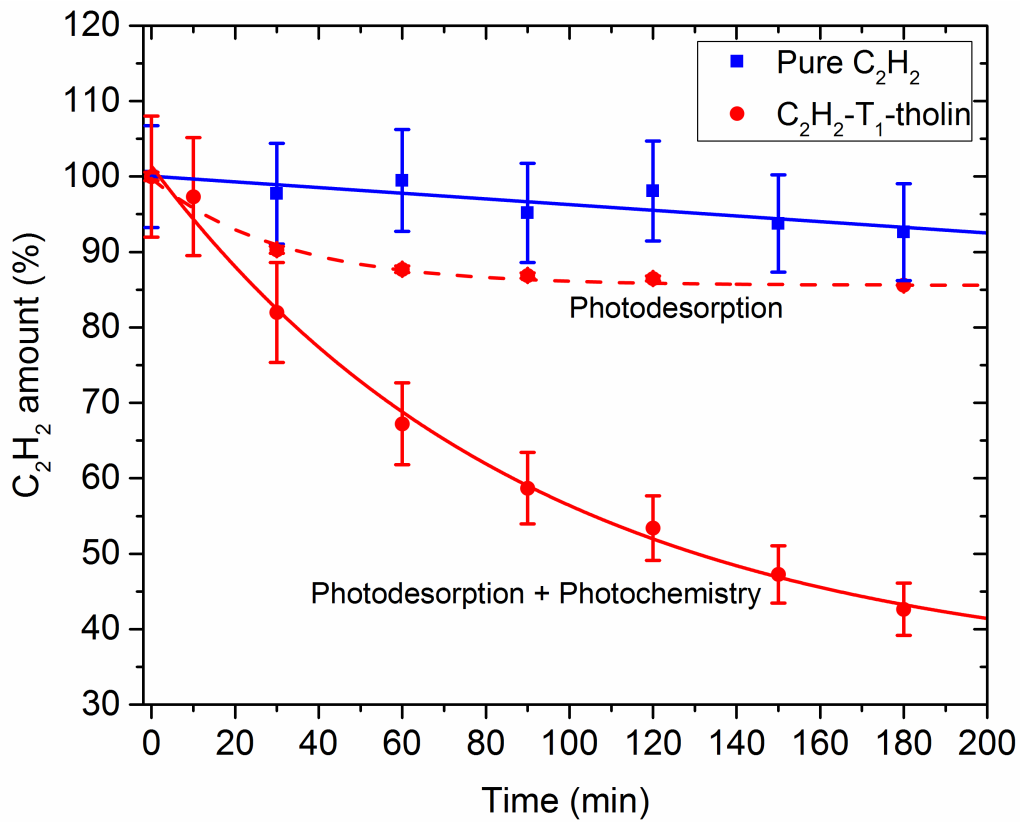

**Figure 6:** Evolution of the $C_2H_2$ amount as a function of the irradiation time measured by infrared spectroscopy (solid line) for pure $C_2H_2$-ice and for $C_2H_2$ deposited on a $T_1$-tholin sample. Evolution of the $C_2H_2$ amount as a function of the irradiation time for the photodesorption process measured by mass spectrometry for the $C_2H_2$-$T_1$-tholin experiment (dash line). The uncertainties are given at 2σ (standard deviation) and were calculated from the standard fluctuations of the infrared spectroscopy and mass spectrometry measurements, respectively.

We found that in the pure $C_2H_2$-ice experiment, the acetylene consumption follows linear kinetics with a total amount of acetylene consumed about 5% after 3 hours of irradiation, while in the $C_2H_2$ on $T_1$-tholin experiment the acetylene consumption follows exponential kinetics with a total amount of acetylene consumed about 60% after 3 hours of irradiation. Figure 6 shows a good correlation for the first 10 minutes of irradiation and then a divergence between these two quantities. Indeed, Figure 6 shows that after 10 minutes of irradiation the total amount of acetylene consumed is still increasing (determined by infrared absorption band integrated intensity at 3225 cm$^{-1}$), while the amount of acetylene photo-desorbed reaches a plateau (measured by QMS). It indicates that the freshly accreted acetylene has higher propensity to photodesorb, whereas aged ice or aerosols desorb far less acetylene. Under our experimental conditions, we found that the acetylene amount consumed by photodesorption was about 15% after 3 hours of irradiation while the total amount of $C_2H_2$ consumed was about 60%. This result highlights that about 25% of the acetylene was lost to photodesorption in these experiments while the other 75% depletion was linked to another process, namely photochemistry.

Since solid acetylene has only a weak $S_0 \rightarrow T_1$ (spin-forbidden) transition at an energy about 3.58 eV (346 nm) (Couturier-Tamburelli et al., 2015; Malsch et al., 2001) and the photon energy used during these experiments was about 3.49 eV (355 nm), Desorption Induced by Electronic Transition (DIET), which has been studied for other ices such as CO (Fayolle et al., 2011), should not be an important process. Further, DIET cannot explain the lack of photodesorption observed in the pure acetylene experiment. We propose an indirect mechanism that could explain the photodesorption of acetylene that occurs selectively when $C_2H_2$ is accreted on $T_1$-tholin but not from the pure $C_2H_2$-ice films. We propose that the photons were primarily absorbed by the $T_1$-tholin and the excitation energy was transferred to acetylene molecules, resulting in electronically and vibrationally excited $C_2H_2$ molecules and thermal desorption of $C_2H_2$ at the surface. Indeed, several studies have highlighted that such a mediated process could occur when one molecule absorbs the photon energy and another molecule desorbs (Bertin et al., 2012; Fillion et al., 2014; Thrower et al., 2008). In our experiments, $T_1$-tholin acted as a very efficient photon absorption and energy transfer system compared to sapphire, which has essentially no photoabsorption at UV-Vis wavelengths. However, a limited acetylene photodesorption mediated by the weak $S_0 \rightarrow T_1$ (spin-forbidden) transition or by the sapphire substrate could occur as also been observed for benzene and water under similar conditions (Thrower et al., 2008). This mechanism could explain the 5% of acetylene depletion in the pure acetylene experiment. Figure 7 presents the absorbance of the $T_1$-tholin and $T_5$-tholin samples in the 250-800 nm range. Tholins do not absorb at λ>500 nm, but the absorption becomes monotonically stronger at wavelengths shorter than 450 nm. In this scenario, the tholin film first absorbs the photons energy and then transfers this energy to the neighboring $C_2H_2$ molecules. Thus, tholins acting as antenna molecules increased the photodesorption efficiency by a factor of 3 at the wavelength considered compared to pure $C_2H_2$ deposited on a sapphire window. Tholin samples are also highly porous (Carrasco et al., 2009) and hence are poor thermal conductors. This property could enable tholins to be electronically excited through the absorption of a photon and convert that energy efficiently into thermal energy, which could be transmitted to acetylene molecules adsorbed in the porous voids, resulting in their photodesorption.

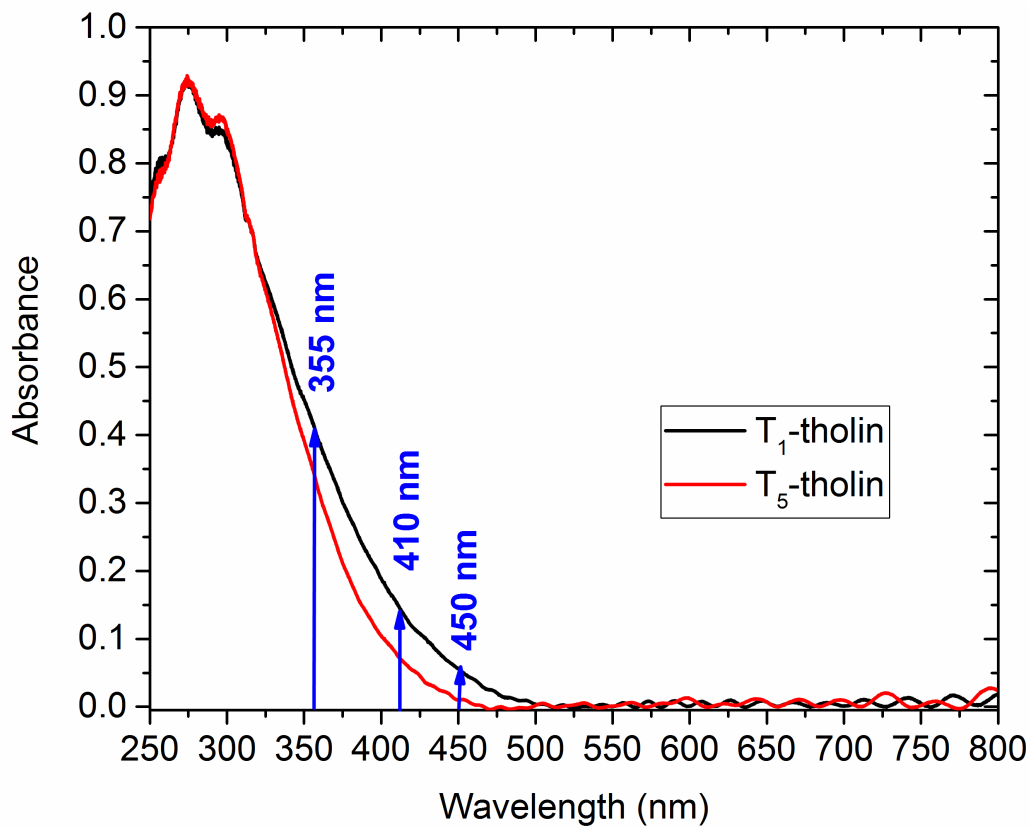

**Figure 7: Absorbance of a $T_1$-tholin sample and a $T_5$-tholin sample (red) in the 250-800 nm range. The blue arrows point to the different wavelengths of irradiation used during the experiments.**

Thus, photodesorption of acetylene mediated by tholins appears to be an important process. However, this process explains only 25% of the acetylene consumption observed by IR spectroscopy. Most likely, the other process, which is responsible for the other 75% of the $C_2H_2$ consumption, should be photochemistry. This possibility is further discussed later in the article.

### 3.1.2. Wavelength dependence of the $C_2H_2$ photodesorption on $T_1$-tholin

It has been shown in other studies that the photodesorption processes are wavelength dependent (Bertin et al., 2013; Fillion et al., 2014; Thrower et al., 2008). In our experiments also photodesorption of acetylene was wavelength dependent but driven by the photoabsorption of $T_1$-tholin. This hypothesis is further corroborated by the dependence of the acetylene consumption from the $C_2H_2$-over-$T_1$-tholin sample as a function of the wavelength of irradiation at 410 nm and 450 nm. To compensate for the lower flux of photons at 410 nm and 450 nm compared to 355 nm as shown in Table 1, we irradiated the sample for longer times at these two wavelengths, 29 hours and 25 hours, respectively, resulting in similar total photon fluence at all three wavelengths (355 nm, 410 nm, and 450 nm). It is important to note here that we have not detected desorption of acetylene using mass spectrometry at 410 nm and 450 nm. However, we have measured a decrease of the IR absorbance of $C_2H_2$ as discussed previously for the 355 nm irradiation. Figure 8 presents the evolution of acetylene concentration, monitored with IR spectroscopy, as a function of the photon-fluence for $C_2H_2$-over-$T_1$-tholin sample for the experiments at 355 nm, 410 nm and 450 nm. Also included here are the total acetylene depletion data (measured by IR spectroscopy) and the depletion data that were due to the photodesorption (measured by mass spectrometry) at 355 nm for comparison.

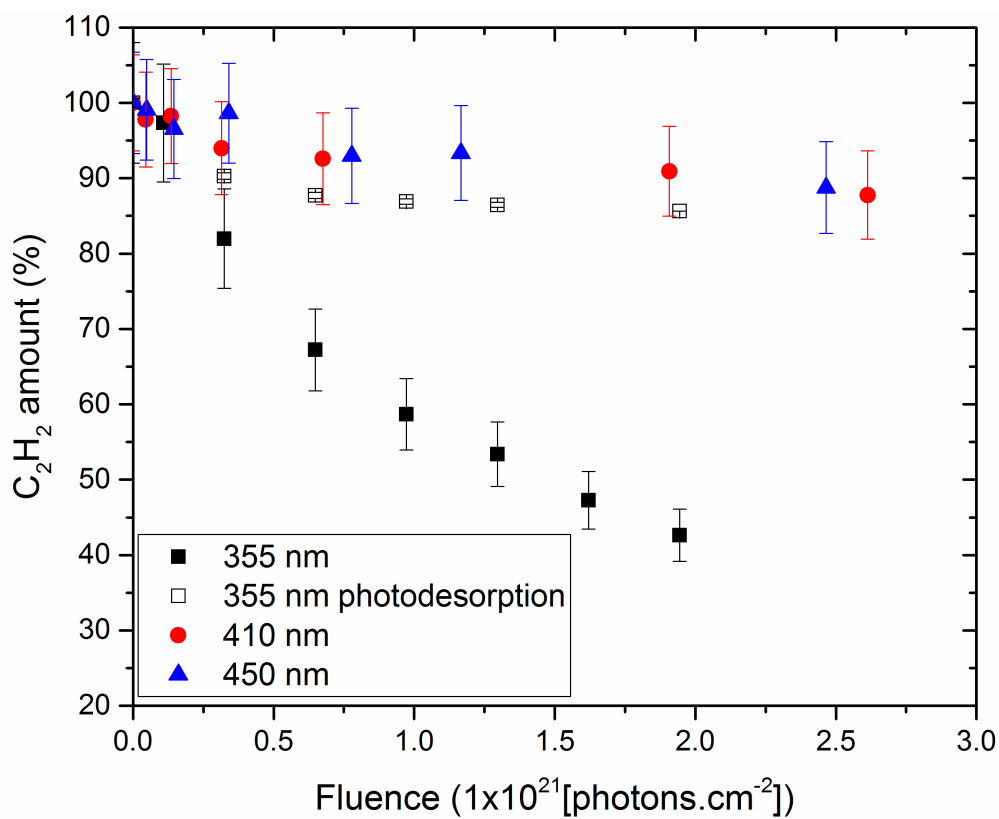

**Figure 8: Evolution of the $C_2H_2$ amount measured by infrared spectroscopy for $C_2H_2$ deposited on a $T_1$-tholin sample as a function of the photon-fluence at 355 nm (black square), 410 nm (red circle) and 450 nm (blue triangle). Photodesorption of $C_2H_2$ at 355 nm was quantified by mass spectrometry whereas the rest of the curves are quantified through infrared absorption, which has much larger error bars in integrated absorbance. The uncertainties are given at 2σ (standard deviation) and were calculated from the standard fluctuations of the infrared spectroscopy and mass spectrometry measurements respectively.**

The total amount of acetylene consumed is similar for 410 nm and 450 nm with a maximum about 10% after an irradiation with a fluence ~$2.5\times10^{21}$ photons·cm$^{-2}$, which corresponds to a total time of irradiation of ~1700 min and ~1500 min respectively according to the photons flux given in Table 1 for these two wavelengths. For both wavelengths, the kinetics is slower and the total amount of acetylene consumed was much lower than the 355 nm irradiations. The UV-visible absorbance spectrum of the $T_1$-tholin presented in Figure 7 shows that tholins still absorb at 410 nm and 450 nm. Therefore, the amount of acetylene consumed measured by IR spectroscopy at these two wavelengths, can be explained by an indirect photodesorption process mediated by the tholins acting as antenna, as discussed earlier. The lack of detection of acetylene by mass spectrometry in the 410 nm and 450 nm experiments may be explained by the detection limit of the mass spectrometer (<$2\times10^{-16}$ A), reached due to the much lower photon flux of the laser used at these two wavelengths compared to 355 nm. In addition, the absorption of $T_1$-tholins is about 3 and 8 times lower at 410 and 450 nm, compared to 355 nm, resulting in 30 to 80 times lower efficiency of energy transfer at these two visible wavelengths. Further, energy transfer from $T_1$-tholin to acetylene becomes even less efficient due to lack of electronic excitations in acetylene at 410 nm and 450 nm, necessitating energy to be transferred through thermal vibrational modes. Hence, the amount of acetylene photodesorbed over the integrated time of 400 milliseconds at these two wavelengths (410 and 450 nm) is below the limit of detection of our mass spectrometer but clearly detectable in IR spectra of the $C_2H_2$-over-$T_1$-tholin sample over the period of time.

### 3.1.3. Influence of the tholins composition on the acetylene reactivity

The results presented in the previous sections highlight that the photoabsorption of tholins at λ<450 nm could mediate photoinduced processes in an accreted $C_2H_2$-ice layer, leading to indirect photodesorption and possibly photochemistry of acetylene. The experimental parameters, such as the energy source, the pressure or the initial gaseous mixture composition, used to simulate Titan's atmospheric chemistry are known to impact the properties of the tholins produced and their ability to reproduce Titan's aerosols properties (Cable et al., 2011), particularly for their optical properties (Brassé et al., 2015). A previous study by Mahjoub et al. (2012) has shown that the optical indices of the tholins produced in our experimental setup were, in particular, sensitive to the percentage of methane used. Tholins produced in a gaseous mixture with a lower amount of $CH_4$ incorporate more nitrogen resulting in a slightly higher absorbance of the tholins in the UV-visible range (Mahjoub et al., 2012). For this reason, we have also studied the impact of the initial composition of the tholins and their optical properties by repeating the experiments that were conducted with $T_1$-tholin, with a tholin sample produced with a higher percentage of $CH_4$, 5% or $T_5$-tholin sample.

First, the comparison of the UV-Vis absorbance spectra of the $T_1$- and the $T_5$-tholin samples shows a similar shape with a strong absorption in the UV starting at 500 nm and increasing for shorter wavelengths. However the $T_1$-tholin sample presents a slightly higher absorbance between 350 nm and 500 nm in agreement with the optical indices measured in a previous work (Mahjoub et al., 2012). Second, we compare the reactivity of $C_2H_2$ deposited on $T_1$- or $T_5$-tholin samples. We found significant differences in the photodepletion of acetylene when accreted over the $T_1$-tholin sample compared to the $T_5$-tholin sample as shown in Figure 9. We observed that the photo-reactivity of acetylene does not reflect the absorption spectra of $T_1$- and $T_5$-tholins (Figure 7), demonstrating that UV-absorption alone was not the determining factor for these differences of reactivity.

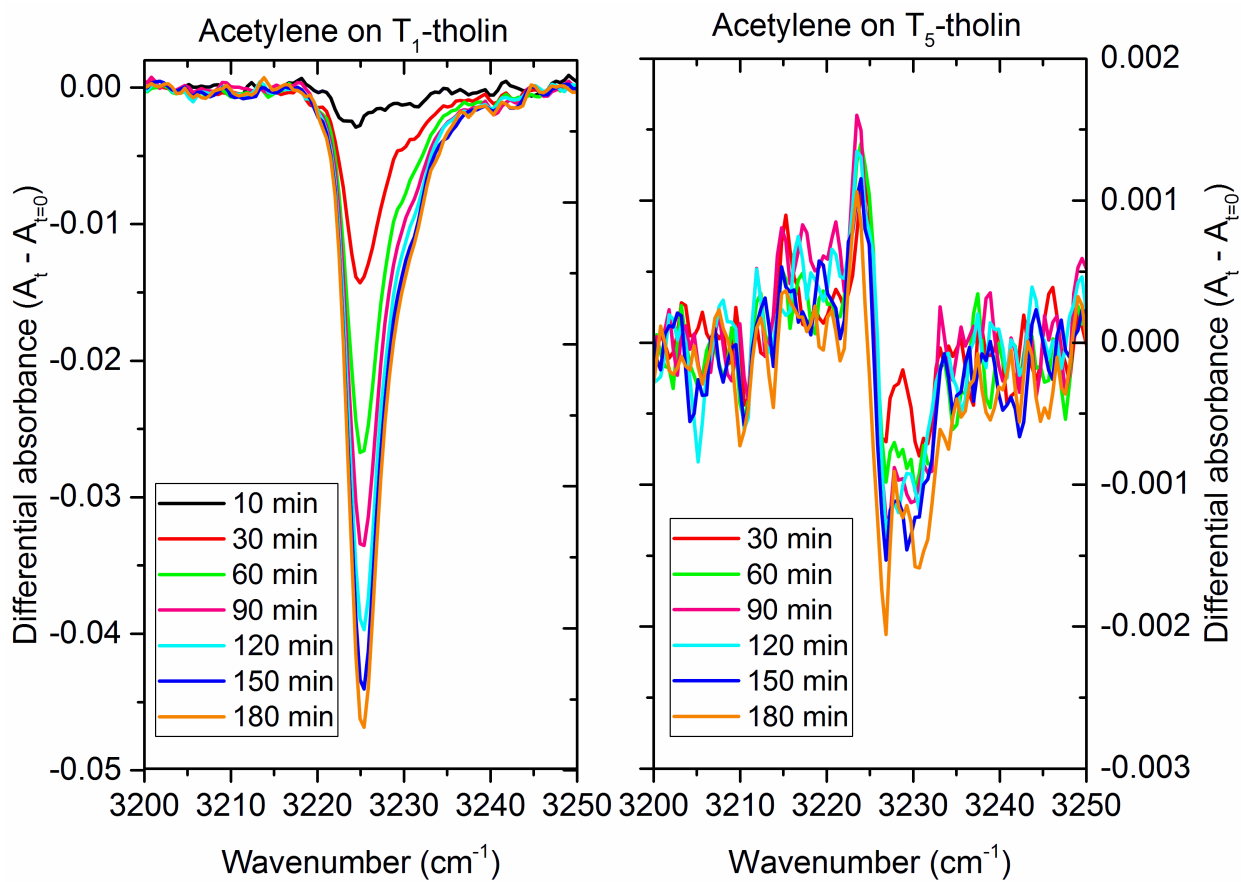

**Figure 9: Evolution of the differential absorbance ($A_t - A_{t=0}$) of $C_2H_2$ as a function of the irradiation duration at 355 nm. The left panel presents results for acetylene coated on a**

**PAMPRE tholin sample produced in a gaseous mixture made of 1% of $CH_4$ in $N_2$ ($C_2H_2$-over-$T_1$-tholin). The right panel presents results for acetylene coated on a PAMPRE tholin sample produced in a gaseous mixture made of 5% of $CH_4$ in $N_2$ ($C_2H_2$-over-$T_5$-tholin).**

Though we detected photodepletion of $C_2H_2$ in the experiments carried out with the $T_5$-tholin samples, photochemistry involving acetylene is found to be inefficient compared to the $T_1$-tholin sample. The total depletion of $C_2H_2$ after three hours of irradiation amounts to 6% for the $T_5$-tholin experiment, an order of magnitude lower than 60% observed for the $T_1$-tholin experiment, though the fluence of the 355 nm laser was same in both experiments. The depletion of acetylene ice on $T_5$-tholin is similar to the result obtained in the pure acetylene experiment or with $T_1$-tholin at 410 nm and 450 nm. We attribute the lack of photochemistry of the $C_2H_2$-over-$T_5$-tholin sample as due to the chemical composition difference between $T_1$- and $T_5$-tholins. Indeed, $T_1$-tholins incorporate more nitrogen than $T_5$-tholins (Gautier et al., 2014; Sciamma-O'Brien et al., 2010), leading to tholins enriched in amines (Gautier et al., 2012) for $T_1$-tholin. Increased absorption between 400 and 500 nm of $T_1$-tholin (Figure 7) also indicates higher "unsaturated covalent bonding" in $T_1$-tholin compared to $T_5$-tholin. This difference of composition could influence the indirect photodesorption as well as photochemistry of the acetylene ice. Indeed, the efficiency of this mechanism depends on the nature of interaction between the ice and the UV absorber as well as the possibility for the absorber (tholin) to transfer energy to acetylene (and eventually lead to chemical bonding between acetylene and tholin). The higher the amine and unsaturated bonding in the tholin sample, the higher would be the probability of formation of covalent bonding between tholin and acetylene mediated by soft-excitations with longer wavelength photons. As a result, the efficiency of photodesorption and photochemistry of $C_2H_2$ mediated by $T_1$-tholin films is very significant and these processes are very likely to occur in Titan's atmosphere and on the surface. We conclude that aerosols produced with lesser $CH_4$ mixing with $N_2$ are a potentially important carriers of photochemical sink of acetylene in Titan's atmosphere.

## 4. Discussions

### 4.1. Photodesorption rates and implications for Titan

We can calculate the photodesorption rate of acetylene for our experimental conditions using the following Eq. (2):

$$Y_{PD} = \frac{N}{\Phi}, \tag{2}$$

where $Y_{PD}$ is the photodesorption rate (molecules·photon$^{-1}$), N is the molecular column density (molecules·cm$^{-2}$) and $\Phi$ is the fluence (photons·cm$^{-2}$). N can be calculated as the product of the initial column density $N_0$ and the amount of acetylene photodesorbed as measured by mass spectrometry. $N_0$ is calculated from the initial absorbance of $C_2H_2$ before the irradiation using the following Eq. (3):

$$N_0 = \frac{\int \tau_v dv}{A}, \tag{3}$$

where $N_0$ is the initial column density (molecules·cm$^{-2}$), $\int \tau_v dv$ is the integral of the band in optical depth, which can be derived from the absorbance value. A is the band-strength (cm·molecule$^{-1}$). For our calculation, we used a band strength value of $3.58 \times 10^{-17}$ cm.molecule$^{-1}$ for the 3225 cm$^{-1}$ band calculated by Hudson et al. (2014) for the crystalline acetylene at 70 K. In our experiments, the $N_0$ values are found to be $2.89 \times 10^{16}$ molecules·cm$^{-2}$. We obtained on an average a desorption rate of $(2.1 \pm 0.2) \times 10^{-6}$ molecules·photon$^{-1}$ averaged on the total irradiation time at 355 nm. This is much lower than the values determined for the photodesorption of CO, $N_2$ or $CO_2$ ices by high-energy photons ($\lambda < 170$ nm), which varies from $1 \times 10^{-3}$ to $4 \times 10^{-2}$ molecules·photon$^{-1}$ (Bertin et al., 2013; Fillion et al., 2014).

The residence time of the aerosols in the atmosphere of Titan has been calculated using both a 1-D microphysical model (Lavvas et al., 2011b; Lavvas et al., 2010) and a general circulation model (GCM) including the microphysics of aerosols (Larson et al., 2014), based on which it was estimated that the total lifetime of the aerosols in the atmosphere to be 395 Titanian years, which corresponds to $\sim 8.9 \times 10^9$ s. For the wavelength of 350 nm, the photon flux in the lower atmosphere of Titan varies from $\sim 3 \times 10^{11}$ photons·cm$^{-2}$·s$^{-1}$ at 200 km of altitude to $\sim 8 \times 10^7$ photons·cm$^{-2}$·s$^{-1}$ at 70 km of altitude (Gudipati et al., 2013; Wilson and Atreya, 2004). According to the model of Larson et al. 2014, aerosols spend most of their time in the lower atmosphere. Therefore, we can use the total lifetime, and the flux at 200 km, which is the higher limit of the photon flux received at this wavelength in the stratosphere and the troposphere of Titan to calculate an upper limit of the fluence received by ices coated on aerosols. We obtained a value of $2.7 \times 10^{21}$ photons·cm$^{-2}$ at 350 nm to be delivered to an aerosol before reaching Titan surface, the same order of magnitude as the one used in our experiment at 355 nm (*i.e.* $1.8 \times 10^{21}$ photons·cm$^{-2}$). It is likely that the photon fluence received by an ice particle in Titan atmosphere would be lower than this upper limit. Indeed, we should keep in mind that the photon flux at 350 nm decreases by several orders of magnitude with the altitude in the lower atmospheric layers as these particles travel from the stratosphere to the troposphere and on to the surface. Further, our data is obtained for a given monochromatic laser wavelength and our data needs to be integrated to represent the range of wavelengths reaching the lower atmosphere and the surface of Titan, which could increase the photodesorption yields – but not photochemistry yields – based on the data presented in Figure 8. Consequently, photodesorption could open up a replenishing channel for a part of the acetylene ice to be converted into the gas-phase and fed into Titan's atmosphere, even if further quantification is necessary to better understand the importance of this process. Thus, it is important to include photodesorption to describe the microphysics of Titan's aerosols, particularly involving highly-abundant and highly-volatile ices such as acetylene. Finally, similar processes could be occurring with other condensed volatiles in Titan's ice particles, aerosols, and on the surface.

### 4.2. Photochemistry of acetylene on tholins

We have demonstrated that tholins could drive photodesorption of coated solid acetylene, while irradiated at UV-Vis wavelengths where the tholins have a strong absorption, but not acetylene itself. In the case of the experiments conducted with the $T_1$-tholin samples and irradiated at 355 nm, the amount of acetylene desorbed, as quantified by mass spectrometry, cannot account for the total amount of acetylene consumed as determined by infrared spectroscopy. Another process has to be responsible for the rest of the acetylene (~75%) consumed in these experiments. Direct solid-state photochemistry of $C_2H_2$ required higher energy photons than the 355 nm used in this experiment, as absorption of the acetylene (for the first excited $S_0$-$S_1$ state) starting at wavelength shorter than 237 nm (Couturier-Tamburelli et al., 2015; Malsch et al., 2001). Moreover, the direct acetylene ice photochemistry in the VUV has been experimentally studied, highlighting the formation of longer hydrocarbons molecules, including the formation of polymer (Compagnini et al., 2009; Cuylle et al., 2014). They were not observed in our experiments, excluding direct photochemistry in the acetylene layer, also in agreement with our observation that no significant depletion of $C_2H_2$ was detected when pure $C_2H_2$ was deposited on a sapphire window. This leaves the possibility for photochemical reactions of acetylene with the tholins further propagating into the crystalline $C_2H_2$ ice layer as shown in a previous experimental study for nitrile compounds (Couturier-Tamburelli et al., 2018). In order to explore this possibility, we analyzed infrared spectra obtained before the beginning of the experiment and subsequent to warming up of the photolyzed $C_2H_2$-$T_1$-tholin film at 300 K, as shown in Figure 10. We should note here that the tholin film thickness is approximately 600 nm, whereas the acetylene film thickness is ~15 – 20 nm, of which 60% photodepleted (15% photodesorption and 45% photochemistry), resulting in ~1% material addition to the tholin at the end of this experiment.

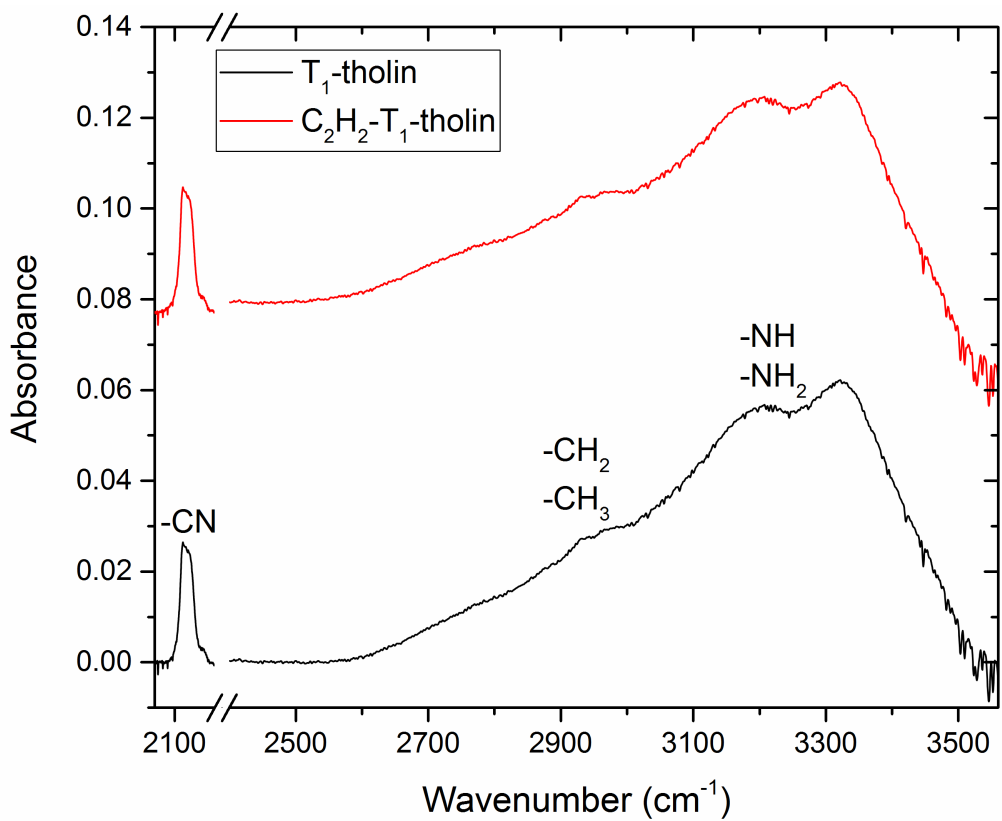

**Figure 10: Infrared spectra of the T1-tholin sample at room temperature before the irradiation (black) and subsequent to warming up the photolyzed $C_2H_2$-$T_1$-tholin (red).**

From the spectra shown in Figure 10, we did not find significant changes in these spectra. It is likely because while acetylene C-H infrared absorption band-strength is very high, when incorporated into the tholin film covalently, the resulting C-H absorption may not be of the same band-strength and that the environment in which such bonding occurred may vary, resulting in very broad new absorbance of the photochemically incorporated $C_2H_2$ in the $T_1$-tholin. Indeed, polymer IR absorptions tend to much broader and the new signals may be buried under the broad polymer absorption. Thus, infrared spectra did not yield convincingly concrete new information on the potential chemical structures of incorporated $C_2H_2$. Our experiments using isotopically labeled acetylene ($^{13}C_2H_2$) also did not yield any outstanding new features in the room-temperature infrared spectrum that could have been assigned to incorporation of $^{13}$C-H into the tholins, again most likely due to broad absorption features and small isotope shifts between $^{12}$C-H and $^{13}$C-H frequencies being merely 10 cm$^{-1}$. However, we observed a slight change in the broad structure around 3100 cm$^{-1}$ between before and after photoreactivity of $C_2H_2$ with $T_1$-tholin such that after the photolysis the vertical gap between the -$CH_2$/-$CH_3$ absorption around 2900 cm$^{-1}$ and the peak for –NH/-$NH_2$ at 3200 cm$^{-1}$ decreased. This could be due to the expected broad absorption due to the incorporated acetylene that could have resulted in $H_2$C=HC-N-(tholin) after reacting with an amine functional group. New experiments that are highly sensitive to surface chemical composition at a scale of a few nanometers are necessary to better understand the reactivity occurring at the interface between the tholins and the ice layer and the possible photochemical modifications in tholins – the laboratory analogs of Titan's atmospheric aerosols.

## Conclusion

We have carried out a study of the photo-induced processes of acetylene ice film coated on laboratory analogues of Titan's aerosols – tholins - at wavelengths longer than 300 nm. We have demonstrated that the aerosols can act as antennae by absorbing the photons in the UV-Vis wavelength regions and efficiently drive the photodepletion of accreted acetylene, 60% of the initial acetylene was depleted after a fluence of $1.8 \times 10^{21}$ photons·cm$^{-2}$ at 355 nm.

Monitoring of the residual gases in the vacuum chamber by mass spectrometry has allowed us to determine that 15% of the initial acetylene was photodesorbed. This is an indirect process driven by the absorption of photons first by tholins at the wavelength used for the irradiation and subsequently transferring that energy to accreted acetylene ice. We have calculated a photodesorption rate of acetylene to be $(2.1 \pm 0.2) \times 10^{-6}$ molecules·photon$^{-1}$ averaged on the total irradiation time at 355 nm. The photon fluences used in our experiment are an upper limit for the one received at 355 nm by the aerosols during their sedimentation through the stratosphere and the troposphere of Titan, where condensed species are observed. Integrated photon flux at different altitudes of Titan's atmosphere would help better quantification of the photodesorption efficiencies. To better constrain the importance of this process, it would be necessary to extend our studies to a range of photon wavelengths (>250 nm), from the altitudes where $C_2H_2$ condenses (~70 km) down to the surface, and then model their fluences at various altitudes in Titan's lower atmosphere.

We found that the photodesorption process alone could not explain all of the acetylene photodepletion that was measured. Since the direct solid-state photochemistry of solid acetylene is not possible at 355 nm, acetylene should react photochemically with the $T_1$-tholins during the energy transfer process, where unsaturated bonds and amine functional groups become easy locations for acetylene reaction with tholins. New and more sensitive surface analysis methods that can interrogate the upper few nanometers are needed to confirm our hypothesis that tholins first absorb photons and transfer this energy to $C_2H_2$ molecules for desorption or initiate chemical bonding through excited-state photochemistry between unsaturated tholin centers and acetylene.

## Acknowledgments

This work is supported by the NASA Solar System Workings grant "Photochemistry in Titan's Lower Atmosphere". The research work has been carried out at the Jet Propulsion Laboratory, California Institute of Technology under a contract with the National Aeronautics and Space Administration. NC thanks the European Research Council for funding via the ERC PrimChem project (grant agreement No. 636829.). We thank Dr. Bryana Henderson for proofreading the manuscript.